**Diversification versus specialization in scientific research: which strategy pays off?[1]**

Giovanni Abramo

*Laboratory for Studies in Research Evaluation
at the Institute for System Analysis and Computer Science (IASI-CNR)
National Research Council of Italy*
ADDRESS: Istituto di Analisi dei Sistemi e Informatica, Consiglio Nazionale delle Ricerche, Via dei Taurini 19, 00185 Roma - ITALY
giovanni.abramo@uniroma2.it

Ciriaco Andrea D'Angelo

*University of Rome "Tor Vergata" - Italy and
Laboratory for Studies in Research Evaluation (IASI-CNR)*
ADDRESS: Dipartimento di Ingegneria dell'Impresa, Università degli Studi di Roma "Tor Vergata", Via del Politecnico 1, 00133 Roma - ITALY
dangelo@dii.uniroma2.it

Flavia Di Costa

*Research Value s.r.l.*
ADDRESS: Research Value, Via Michelangelo Tilli 39, 00156 Roma - ITALY
flavia.dicosta@gmail.com

**Abstract**

The current work addresses a theme previously unexplored in the literature: that of whether the results arising from research activity in fields other than the scientist's primary field have greater value than the others. Operationally, the authors proceed by identifying: the scientific production of each researcher under observation; field classification of the publications; the field containing the greatest number of the researcher's publications; attribution of value of each publication. The results show that diversification at the aggregate level does not pay off, although there are some exceptions at the level of individual disciplines. The implications at policy level are notable. Since the incentive systems of research organizations are based on the impact of scientific output, the scientists concerned could resist engaging in multidisciplinary projects.

# 1. Introduction

The current work addresses a question that has not yet been dealt with in studies on research diversification: what strategy yields the best results in terms of impact on future research advancement – field diversification or specialization (concentration)? The answer could vary depending on the scientists' primary fields of research, and for this, any relevant analysis must be conducted at the field level.

In recent decades there has been remarkable growth in collaboration for purposes of scientific research. Evidence of this has repeatedly been presented in analyses of single and co-authorship (Melin & Persson, 1996), which indicate that the share of single-authored publications is constantly in decline (Abt, 2007; Uddin, Hossain, Abbasi, & Rasmussen, 2012), while average number of authors per publication is continuously increasing across different scientific fields (Persson, Glänzel, & Danell, 2004; Archibugi & Coco, 2004; Wuchty, Jones, & Uzzi, 2007; Gazni, Sugimoto, & Didegah, 2012; Larivière, Gingras, Sugimoto, & Tsou, 2015).

Collaborations can be among individuals working in the same or different fields. The search for solutions to increasingly complex societal problems often requires integration of theories, methods and instruments from diverse disciplines, and indeed a growing share of collaborations are multidisciplinary in nature. Such arrangements are in part stimulated by policy and research-management initiatives (Cassi, Champeimont, Mescheba, & de Turckheim, 2017; Van Rijnsoever & Hessels, 2011). The increasing importance of multidisciplinary collaborations is further evidenced in terms of the number of studies intended to provide relevant indicators of measurement (Wagner et al., 2011; Abramo, D'Angelo, & Di Costa, 2016).

Today's researcher is constantly subject to opposing forces: centripetal, pushing them ever deeper in investigation within their own field of specialization, and centrifugal, drawing them to apply their expertise in multidisciplinary projects, or in fields new to them. For Schuitema and Sintov (2017), the researcher's point of departure is always their own disciplinary competences, while interdisciplinary incursions represent an opportunity to acquire and offer complementary competences. More generally, Chakraborty, Tammana, Ganguly, & Mukherjee (2015) identify two behaviors or attitudes on the part of researchers: leading to scientific activity distributed in different fields, or to strong concentration in just a few fields. These authors observe that the large part of prominent researchers tend to pursue a characteristic "scatter-gather" strategy, while essentially remaining focused on one or two fields over the course of their career. Abramo, D'Angelo, and Di Costa (2017) address related questions in investigating whether and to what extent scientists tend to diversify their research activity, and whether such tendency varies across disciplinary areas. The investigative procedure analyzed the nature of diversification along three dimensions: extent of diversification, intensity of diversification, and degree of relatedness of fields in which researchers diversify.

In activities such as finance and corporate management, "diversification" is often considered as a hedge against risk (Lichtenthaler, 2005). In contrast, according to Bateman and Hess (2015), scientists can view research diversification as unattractive, precisely because they perceive it as risky and relatively unproductive. Indeed, if the utility gained by research activity is a function of its impact, one could suppose that a rationally thinking scientist would generally prefer to concentrate on their specific field, moderated by their perceptions of risk and difficulty in achieving results in domains not centered on their core competence. The degree of diversification could also be partially explained by the personal attitudes and preferences of the scientist, which would function as inhibitors or facilitators for boundary-crossing research. The origin and course of evolution of aptitudes and interests would influence the scope of the individual's later research activities, the criteria for choosing and designing projects, and matters of team participation. In making such choices concerning diversification, researchers can also be influenced by the "recognition" factor (Foster, Rzhetsky, & Evans, 2015). And, returning to the important matters of policy and management strategies, the related monetary incentives are clearly designed to encourage scientists to favour certain directions and choices (Stephan, 2012).

The next section explains the methodology and dataset used for the analysis, while Section 3 presents the results. The paper ends with the authors' conclusions and comments.

## 2. Methodology and dataset

The analysis applies bibliometric techniques for the identification of research outputs, their field classification, and the measure of their value (impact). The critical element of the methodology is the identification of the scientist's core research field, thus enabling comparison of the impact of results from research in that field with the impact of their output in other fields.

Is the scientist's core field the one which contains their most representative work? And then, as corollary: is this most representative work the one work that is most cited?

Ioannidis, Boyack, Small, Sorensen, and Klavans (2014) assert that the vast majority of elite scientists feel their most important paper is indeed among those that are most cited. This conclusion was based on a preceding work, where the authors surveyed the 400 biomedical scientists most cited over the 1996–2011 period, asking their opinions as to their most influential published works (Boyack, Klavans, Sorensen, & Ioannidis, 2013).

But, according to Niu, Zhou, Zeng, Fan, and Di (2016), an author's most-cited work is not necessarily the most representative of their scientific production. Instead, to identify this most representative work, they propose an approach called "self-avoiding potential diffusion process" (SPD), based on the citation graph of the author's entire scientific production. However, any process arriving at a single paper as "representative" for the individual seems exposed to severe risk. Taking an obvious example, relating to the field of bibliometrics, there is the famous case of the physicist Jorge Hirsch. His most-cited work is the one where he introduces the $h$-index of scientific performance (Hirsch, 2005). This would indeed seem to identify bibliometrics as his core field, when in fact the vast majority of his works are in the field of physics, and bibliometrics was a field of diversification. Given such considerations, the most appropriate approach seems to be to define the scientist's core research field as that containing the greatest number of their publications – which is the method used for the current study.

The operational steps of the study are as follows: i) definition of the area of investigation; ii) identification of the scientific production of each researcher under observation over a period of time; iii) field classification of the publications; iv) identification of the "core field", containing the greatest number of the researcher's publications; v) attribution of a value to each publication.

Pursuant to point iv): for simplicity, works issued in the core field are referred to as "specialized publications"; those in all other fields are "diversified publications".

The area of observation is "all Italian professors (full, associate, assistant)". As well as being large in scale, this particular population offers advantages of precision and reliability.

For Italian professors, it is possible to disambiguate the true identity of each individual author. Using an authors' name disambiguation algorithm, each publication (article, review article and conference proceeding) can be attributed to the professor that produced it (D'Angelo, Giuffrida, & Abramo, 2011). The harmonic average of precision and recall (F-measure) of authorships, as disambiguated by the algorithm, is around 97% (2% margin of error, 98% confidence interval).

The publication period chosen is 2004–2008, with citations counted at the close of 2015: a citation window wide enough to achieve accurate impact measurement (Abramo, Cicero, & D'Angelo, 2011). A five-year publication period also reduces problems of paucity of publication within the individual fields, and effects from year-dependent fluctuations (Abramo, D'Angelo, & Cicero, 2012). The analysis is restricted to those fields where the prevalent form of encoding new knowledge is publication in scientific journals. For these, the coverage of bibliometric databases such as Web of Science (WoS) can be variable, but is acceptably representative of the overall production of scientists. Furthermore, for the fields of this context, citation-based indicators can be applied to measure the impact of publications. For brevity, these fields are called the Sciences, distinguished from Social Sciences, Arts & Humanities, which do not meet the standards described above.

The Italian university system offers a further advantage, in that all professors are classified in one and only one field, named "scientific disciplinary sector" (SDS; of which 370 in all). The SDSs are grouped into disciplines, named "university disciplinary areas" (UDAs; 14 in all). The Sciences consist of 9 UDAs (Mathematics and computer sciences, Physics, Chemistry, Earth sciences, Biology, Medicine, Agricultural and veterinary sciences, Civil engineering, Industrial and information engineering) and 192 SDSs. The Italian Ministry of Education, University and Research, maintains a database which indexes all Italian professors and provides such information as their affiliation, field classification, academic rank, and gender.

The classification of each publication is derived from the WoS subject category (SC) of the hosting journal. The scientific production of a given author will typically involve a number of SCs, but the SC containing the greater part of their production is generally expected to coincide with their official field designation (SDS).

As an example of the methodology, Table 1 shows the data for the production of a professor in the Experimental Physics SDS (FIS/01). Over the five-year period, their scientific production amounts to 58 publications. From identification of the SCs associated with the journals hosting these works, it can be seen that the journals most frequently hosting his works are in "Optics", which relates to 33 works either alone or in combination with other SCs. It can thus be concluded that the author's scientific production is 58 publications, of which 33 specialized and 25 diversified. Following this same procedure, it is possible to distinguish the specialized and diversified publications for all scientists.

The question is whether specialization or diversification would give the individual the best results in terms of impact. For this, two indicators are applied, representing complementary views of "impact". These are: i) the median of the impact frequency distribution; ii) the most-cited publication. For each indicator, the question is: are the results greater for the individual's specialized publication, or for their diversified publication?

To control for the differing citation behavior across fields and year of publication, citations received by the publication are field-normalized to the average of the citations for all cited Italian publications of the same year and subject category (Abramo, Cicero, and D'Angelo, 2012), thus arriving at the field-normalized impact. The following uses the generic term "impact", considering both this methodology and the above discussion.

*Table 1: Scientific production of an Experimental physics professor (FIS/01)*

| Journal | No. of publications | Subject categories of the journal | Specialized |
|---|---|---|---|
| Journal of Biomedical Optics | 8 | Biochemical Research Methods; Optics; Radiology, Nuclear Medicine & Medical Imaging | Yes |
| Applied Optics | 6 | Optics | Yes |
| Optics Express | 6 | Optics | Yes |
| Physics In Medicine and Biology | 5 | Engineering, Biomedical; Radiology, Nuclear Medicine & Medical Imaging | No |
| Optics Letters | 4 | Optics | Yes |
| Postharvest Biology and Technology | 3 | Agronomy; Food Science & Technology; Horticulture | No |
| Review of Scientific Instruments | 3 | Instruments & Instrumentation; Physics, Applied | No |
| Applied Spectroscopy | 2 | Instruments & Instrumentation; Spectroscopy | No |
| Opto-Electronics Review | 2 | Engineering, Electrical & Electronic; Optics; Physics, Applied | Yes |
| Photochemistry and Photobiology | 2 | Biochemistry & Molecular Biology; Biophysics | No |
| Physical Review Letters | 2 | Physics, Multidisciplinary | No |
| Advanced Photon Counting Techniques II | 1 | Biophysics; Remote Sensing; Optics | Yes |
| Analytical And Bioanalytical Chemistry | 1 | Biochemical Research Methods; Chemistry, Analytical | No |
| Applied Engineering In Agriculture | 1 | Agricultural Engineering | No |
| Biomedical Applications of Light Scattering II | 1 | Engineering, Biomedical; Optics | Yes |
| Biosystems Engineering | 1 | Agricultural Engineering; Agriculture, Multidisciplinary | No |
| Diagnostic Optical Spectroscopy in Biomedicine IV | 1 | Optics; Radiology, Nuclear Medicine & Medical Imaging; Spectroscopy | Yes |
| IEEE Transactions on Instrumentation and Measurement | 1 | Engineering, Electrical & Electronic; Instruments & Instrumentation | No |
| Journal of Applied Physics | 1 | Physics, Applied | No |
| Journal of Texture Studies | 1 | Food Science & Technology | No |
| Molecular Imaging | 1 | Biophysics; Optics | Yes |
| Novel Optical Instrumentation for Biomedical Applications III | 1 | Optics; Physics, Applied; Radiology, Nuclear Medicine & Medical Imaging | Yes |
| O3a: Optics for arts, Architecture, and Archaeology | 1 | Archaeology; Art; Optics; Imaging Science & Photographic Technology | Yes |
| Optical Tomography and Spectroscopy of Tissue VI | 1 | Engineering, Biomedical; Instruments & Instrumentation; Optics; Radiology, Nuclear Medicine & Medical Imaging; Spectroscopy | Yes |
| Physical Review E | 1 | Physics, Fluids & Plasmas; Physics, Mathematical | No |
| Technology in Cancer Research & Treatment | 1 | Oncology | No |

The dataset is composed of the Italian professors in the 192 Science SDSs with 2004-2008 scientific production meeting certain requirements, applied for reasons of significance. These are, that the total production:
- Is composed of at least 5 publications.
- Falls in at least 2 different SCs.

- Presents a single prevalent SC.

The dataset thus prepared consists of 17,698 professors, distributed among the Sciences as follows (per UDA, Table 2).

*Table 2: Dataset of the analysis*

| UDA | SDSs | Professors | Publications | Authorship |
|---|---|---|---|---|
| Mathematics and computer science | 10 | 1,142 | 11,213 | 14,240 |
| Physics | 8 | 1,768 | 21,977 | 56,536 |
| Chemistry | 12 | 2,348 | 23,448 | 46,137 |
| Earth sciences | 12 | 482 | 3,962 | 5,498 |
| Biology | 19 | 2,695 | 24,925 | 40,760 |
| Medicine | 50 | 5,328 | 53,461 | 106,387 |
| Agricultural and veterinary sciences | 30 | 1,190 | 8,812 | 15,401 |
| Civil engineering | 9 | 325 | 2,948 | 3,680 |
| Industrial and information engineering | 42 | 2,420 | 29,741 | 44,302 |
| Total | 192 | 17,698 | 162,918 | 332,941 |

## 3. Analysis and results

Figure 1 charts the numbers of specialized and diversified publications for each of the 17,698 professors considered. The diagram reveals substantial correlation (Pearson $\rho$ equal to 0.70) between the two distributions. This is as expected, since both values depend on the intensity of the professor's production.

Taking the median of the two distributions (8 for number of specialized publications; 4 for number of diversified), four quadrants are obtained. The highest numerosity of professors (5,587 individuals, 31.6% of total) is registered in the upper right quadrant: i.e. showing individuals with both types of publication above median. Next is the lower left quadrant, with 4,525 professors (around one quarter of total): these are individuals with both types of publication below median. The lower right quadrant (diversified above median, specialized below median) registers 2,665 professors (15.1%): evidently individuals with highly diversified production. The contrary case of highly specialized production (upper left) shows 2,082 observations (11.8% of total).

(The remainder of 2,839 observations, or 16.0%, concerns professors with a number of publications exactly equal to one or both medians.)

As with all bibliometric dimensions, the impact measured for the elements of a publication portfolio shows three typical traits:
- Highly skewed distribution;
- Significant number of nil values; and
- Presence of maximum outlier values which impact significantly on the average.

All this suggests the use of the median as index of central tendency and non-parametric test to answer the research questions.

For 9,050 professors (51.1% of total) the median citation impact for their personal publication portfolio is higher for specialized publications. The coefficient of correlation (Spearman's $\rho$) between the median for the two sets of publications is 0.313, and statistically significant. This indicates that the two sets of publications are not independent in terms of impact; there is a monotonic association between the two variables in the population; or in other words, as one might expect, professors with high median impact for specialized publications also show a high median impact for diversified publications.

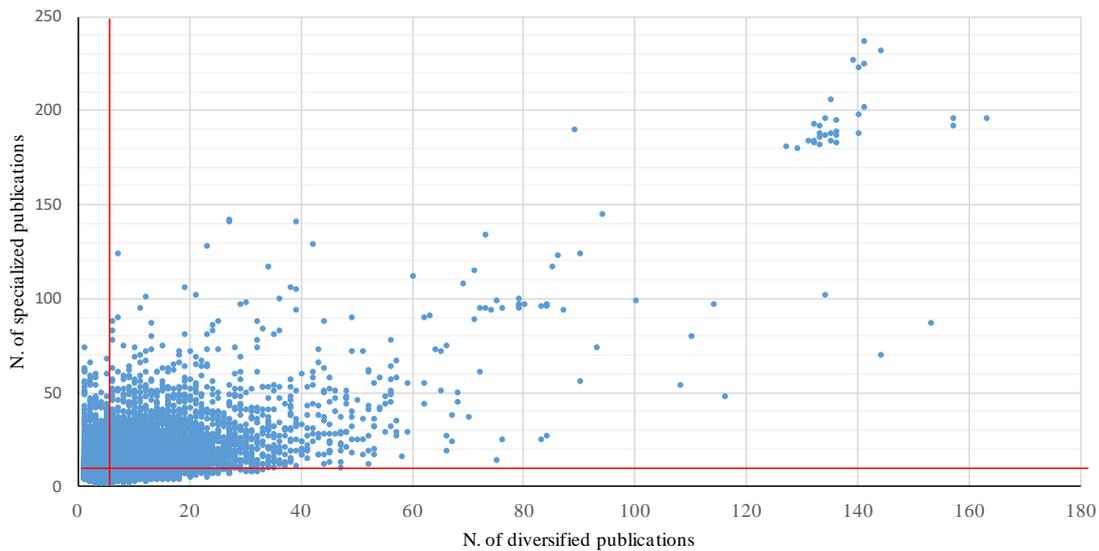

*Figure 1: Dispersion diagram of the number of specialized vs diversified publications for Italian academics (2004-2008 production)*

Figure 2 presents the median citation impact for the specialized and diversified publications of each of the 17,698 professors considered. Almost all professors place in the interval 0-5, for both values of median impact. The figure reveals two types of "extreme" combinations: i) high median citation impact for specialized publications, vs low median for diversified (0.1% professors); ii) low median citation impact for specialized publications vs high for diversified (0.6% professors).

The extreme combinations raise interesting questions, precisely concerning strategy. Some reasons for such combinations can be suggested. Considering that the individual often enters into research diversification through participation in multidisciplinary research projects, this context, including the quality of the other project participants, would be expected to affect (positively or negatively) the impact of their coauthored output. The individual's choice between specialized and diversified research (publication) can then lead to exceptionally different results in terms of impact. Another consideration is that there could be an initial mismatch between the distinctive competencies of the researcher and those required as they attempt to enter a specific new field, again resulting in the extreme combination of high specialized/low diversified medians.

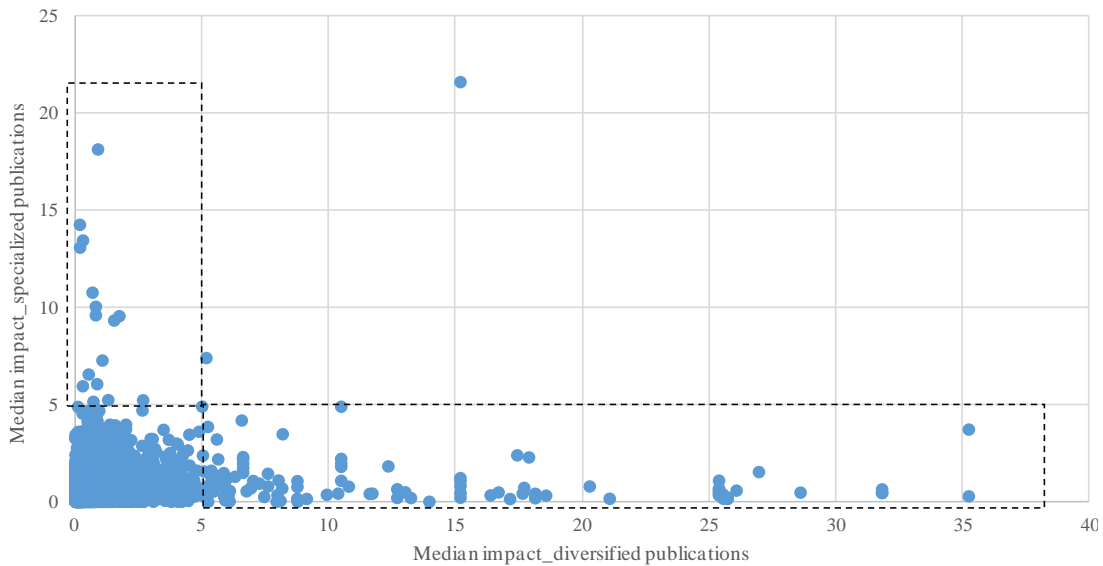

*Figure 2: Dispersion diagram of the median impact of specialized vs diversified publications for Italian professors (2004-2008 production)*

The Wilcoxon signed-rank test is used to verify the null hypothesis: that there are not substantial differences between the two distributions of data (meaning between the impact medians of specialized and diversified publications). The test results (p-value = 0.0282) indicate rejection of the hypothesis: there is a statistically significant difference between the two distributions. Table 3 reports the Wilcoxon matched-pairs signed-ranks test results and Spearman ρ coefficients for the two distributions by UDA. Spearman ρ values vary from a minimum 0.148 in Civil engineering to maximum 0.293 in Biology, and are statistically significant in all UDAs. There are significant differences between specialized and diversified categories in five UDAs out of nine: Physics, Chemistry, Medicine, Agricultural and veterinary sciences, Industrial and information engineering. Observing the incidence of number of professors with median impact of specialized publications higher than for diversified publications, this varies from a maximum of 57.1% in Chemistry to a minimum of 46.1% in Industrial and information engineering (together with Biology, the only UDAs with values below 50%).

In general, it might be expected that a professor would express superior capabilities in their core field, and that the stock and level of accumulated knowledge in one's core field would generate greater impact on the scientific community of reference, compared to their production in other fields. Yet in three UDAs (Biology; Agricultural and veterinary sciences; Industrial and information engineering) this hypothesis does not seem verified.

*Table 3: Spearman's ρ and Wilcoxon matched-pairs signed-ranks, by UDA*

| UDA* | Professors | No. of professors where Hp 1 holds true[†] | Spearman ρ | Wilcoxon matched-pairs signed-ranks test (Prob > \|z\|) |
|---|---|---|---|---|
| 1 | 1,142 | 588 (51.5%) | 0.284 | 0.634 |
| 2 | 1,768 | 963 (54.5%) | 0.222 | 0.012 |
| 3 | 2,348 | 1,341 (57.1%) | 0.264 | 0.000 |
| 4 | 482 | 258 (53.5%) | 0.276 | 0.231 |
| 5 | 2,695 | 1,314 (48.8%) | 0.293 | 0.057 |
| 6 | 5,328 | 2,737 (51.4%) | 0.249 | 0.040 |
| 7 | 1,190 | 559 (47.0%) | 0.162 | 0.005 |
| 8 | 325 | 174 (53.5%) | 0.148 | 0.464 |
| 9 | 2,420 | 1,116 (46.1%) | 0.283 | 0.015 |

\* 1, Mathematics and computer science; 2, Physics; 3, Chemistry; 4, Earth sciences; 5, Biology; 6, Medicine; 7, Agricultural and veterinary sciences; 8, Civil engineering; 9, Industrial and information engineering

[†] Hp 1: Median impact of specialized publications is higher than median impact of diversified publications

A further analysis serves to identify the most cited article (MCA) of each professor, and to reveal whether these fall in the "specialized" or "diversified" category (35 professors hold more than one MCA). Table 4 provides the results, showing that for 61.4% of professors, the MCA is achieved from research in the individual's core field, while for 38.6% the MCA is achieved from diversified research. By UDA, the data show that the percentage of professors with MCA in the specialized category is always greater than 50%: values for this range from a minimum of 52.7% in the Biology UDA to a maximum of 70.5% in Mathematics.

In terms of median impact, as analyzed above, the benefits of specialization were not so striking, at either the aggregate level or in those few individual UDAs where in fact advantages could be observed. In contrast, the analysis of production of MCAs indeed reveals remarkable advantages from specialization.

*Table 4: Classification of "most cited articles" by each professor as diversified or specialized, per UDA*

| UDA* | Professors MCA specialized | MCA diversified | % MCA specialized |
|---|---|---|---|
| 1 | 799 | 334 | 70.5% |
| 2 | 991 | 769 | 56.3% |
| 3 | 1,447 | 899 | 61.7% |
| 4 | 295 | 186 | 61.3% |
| 5 | 1,419 | 1,275 | 52.7% |
| 6 | 3,271 | 2,048 | 61.5% |
| 7 | 739 | 448 | 62.3% |
| 8 | 211 | 112 | 65.3% |
| 9 | 1,659 | 745 | 69.0% |
| Total | 10,831 | 6,816 | 61.4% |

\* 1, Mathematics and computer science; 2, Physics; 3, Chemistry; 4, Earth sciences; 5, Biology; 6, Medicine; 7, Agricultural and veterinary sciences; 8, Civil engineering; 9, Industrial and information engineering

# 4. Discussion and conclusions

## 4.1 Implications for theory

The current study addresses a question previously unexplored in the literature: given the scientific outputs of individual scientists, within their primary research field and "outside", does one of these kinds of production have greater impact? In other words: does research diversification at the individual level pay off?

The results seem to indicate that research outputs in a field other than the scientist's prevalent one achieve lower impact on scientific advancement than those produced in their own field.

The analysis was conducted on the totality of Italian professors operating in the Sciences, observing their production indexed in the WoS over 2004-2008, and applying two indicators of impact. A first analysis verified whether the median citation impact of the each professor's publications was greater or lesser for those in their core research field or for those in other fields. A second analysis verified whether the most-cited publication of the individual professor arose from research within or outside their core field.

At the aggregate level, the results show a statistically significant difference between the medians of impact for specialized and diversified publications, in favor of specialized works. The same analysis conducted at the level of individual UDAs reveals that median impact of diversified publications is higher in only three UDAs (Biology, Agricultural and veterinary sciences; Industrial and information engineering); however it should be noted that the two distributions of median impact are significantly different in five UDAs out of nine (Chemistry; Physics; Agricultural and veterinary sciences; Industrial and information engineering; Medicine).

In terms of median citation impact, the advantages of specialization are not so remarkable, either at aggregate level or where favorable results can be observed in the individual UDAs. However, striking advantages appear when examining most-cited publications: in all the UDAs, the scientists' most-cited publications are prevalently from their core fields of research.

Alternatively, multidisciplinary research might be aimed at solving more complex problems, therefore results are more rarely cutting-edge; or simply, being research more likely applied than basic, it tends to be less cited by scientists. Finally, multidisciplinary research might explore new fields, with a lower number of scientists (potential citers) interested in them.

Although not easy, finding the explanations for these observed phenomena would clearly be of interest. Some of the diversified output from any population of scientists would clearly be the outcome of engagement in multidisciplinary projects, with team members from different fields. A possibility is that such teams are less efficient, leading to results of lesser value. Alternatively, a tendency could be that most multidisciplinary research is not at the cutting edge, but instead aimed at solving what remain as complex problems, or in developing applications of existing results. Both cases might be expected to result in fewer citations for the research products. Finally, multidisciplinary research might be undertaken for exploration of new fields, with the results being of interest to a lower number of scientists, particularly if the work was assumed as a personal initiative.

## 4.2 Implications for policy

The policy implications of these findings are notable. Higher education institutions constitute an important pillar of national research and innovation systems, and the policy agendas of many countries therefore place high priority on strengthening both the institutions themselves and their links with industry. One expression of this policy strategy is the increasing diffusion of national research assessment exercises. The original British RAE served as inspiration, and assessment exercises are now becoming regular events in different nations. This is particularly the case where there is a desire to introduce new principles of governance, or more innovative management in the research sphere. Performance-based research funding systems (PBRFs) are a common resort, and these in turn depend on the results of the national assessments. PBRFs are adopted with the intention of achieving improvement in underperforming institutions (Herbst, 2007, p. 90), and for stimulation of continuous improvement in productivity over the whole research system (Abramo, D'Angelo, & Caprasecca, 2009). An international comparative analysis on the adoption of PBRFs, by Hicks (2012), indicates that subsequent to commencement of the RAE, at least 14 other countries (11 in the EU, China, Australia, New Zealand) have chosen to use the results from national assessment exercises as the basis for the award of some portion of public financing to research institutions. The planning of these exercises has always involved choices between two methodologies: peer review and bibliometrics. Until recently, the most commonly adopted method was peer review, in which products submitted by institutions are evaluated by panels of appointed experts. But as advances have been made in bibliometric methodologies and indicators, many governments have chosen to introduce the use of such metrics, either integrated with peer review in full substitution. Just as in this work, these national exercises are then using citations as a proxy for the value or impact of articles.

At the next lower level, a growing number of universities and research institutions now also structure their internal incentive systems, recruitment and promotion procedures, around research performance: i.e. quantity and impact of scientific output, once again measured in full or in part through bibliometrics. These institutional strategies are often both an extension and a desired consequence of national polices and strategies of PBRF.

However, the current research reveals a risk that scientists, both expecting and then observing lower citation impacts from diversified output, could resist engagement in research projects outside their core fields of research.

It has been shown that diversification behavior is affected by gender, age, and academic rank of researchers (Abramo, D'Angelo, & Di Costa, 2017). The organizational, economic and societal utility of solving complex problems of multidisciplinary character could conflict with the individual utility of achieving adequate recognition and reward for one's own research activity. To counterbalance forces against research diversification, provision should then be made for specific incentives to foster multidisciplinary research collaborations, or for assessment of individual performance that can account for the less visible research arising from multidisciplinary commitments. Specific scientometric techniques have been recently developed to measure and study the extent, intensity, and relatedness of field diversification and interdisciplinarity in research (Wagner et al., 2011; Abramo, D'Angelo, & Di Costa, 2017).

## 4.3 Implications for practice

Research-based companies engage in the search for solutions to complex problems, of multidisciplinary character. Nations increasing rely on policies intended to promote university-industry links, for the achievement of such solutions and the expected economic and social benefits. Any resistance of professors regarding engagement in multidisciplinary/applied research projects would negatively affect public-private research collaboration and contract research.

In this regard, it should be noted that within the academic sphere in particular, there have already been alarms raised concerning the potential of over-emphasis on public-private multidisciplinary and applied research. The lowered involvement of professors in basic research, as a result of favoring applied R&D activities, is argued as counterproductive in the mid and long term (Manjarres, Gutierrez-Gracia, & Vega-Jurado, 2007). Indeed, the productivity of most-performing scientists seem to decrease when involved in long-term relationships with one specific industry-related sponsor (Goldfarb, 2008). While academic-industry collaboration can enhance cross-sector contact between researchers, the secrecy rules that such collaboration entails tend to restrict communication among academics. Company efforts to patent and commercialise research results can conflict with the interests of individual professors in publishing and achieving citations (Welsh, Glenna, Lacy, & Biscotti 2008). A number of studies and program evaluations have indicated that too much industry influence on academic research can undermine future pay-offs from academic research, due to the observed effects of distracting researchers from basic and curiosity-driven research in their specialized fields, and because of the obstruction of traditional values of academic freedom (Berman, 1990, Dosi, Llerena, & Labini, 2006, Goldfarb, 2008).

The findings of the current study would seem to present further observable justification of the hesitation for academics to enter into collaboration involving the private sector and multidisciplinary studies. (Once again, policies and strategies would have to take these effects into account.)

Recently, Abramo, D'Angelo, and Di Costa (2018) investigated the effect of multidisciplinary collaboration on research diversification. They found that an academic's outputs resulting from research diversification are more often than not the result of collaborations with multidisciplinary teams. The effect becomes more pronounced with larger and particularly with more diversified teams. The above results concerning research collaboration and diversification in academic research differ from what is observed in industry, where it seems there is no strong relationship between the occurrence of multi-authored, cooperative R&D and any diversification in the R&D itself (Miyata, 1996).